# Superconductivity in CVD Diamond Thin Film Well-Above Liquid Helium Temperature

Yoshihiko Takano[1*], Masanori Nagao[1], Kensaku Kobayashi[2],
Hitoshi Umezawa[2], Isao Sakaguchi[1], Minoru Tachiki[1], Takeshi Hatano[1],
Hiroshi Kawarada[2]

[1] *National Institute for Materials Science, 1-2-1 Sengen, Tsukuba 305-0047 Japan.*
[2] *School of Science and Engineering, Waseda University, 3-4-1 Okubo, Shinjuku, Tokyo 169-8555, Japan.*

We report unambiguous evidence for superconductivity in a heavily boron-doped diamond thin film deposited by microwave plasma assisted chemical vapor deposition (MPCVD) method. An advantage of the MPCVD deposited diamond is that it can contain boron at high concentration, especially in (111) oriented films. Superconducting transition temperatures are determined by transport measurements to be 7.4K for Tc onset and 4.2K for zero resistance. And the upper critical field is estimated to be 7T. Magnetization as a function of magnetic fields shows typical type two superconducting properties.

*To whom correspondence should be addressed,
Name: Yoshihiko Takano
Postal address: 1-2-1, Sengen, Tsukuba 305-0047 Japan
Phone number: +81-29-859-2842
Fax number: +81-29-859-2801
Email: TAKANO.Yoshihiko@nims.go.jp

Diamond has always been adored as a jewel.  What are even more fascinating about diamond are its outstanding physical properties; it is the hardest material ever known in the world with the highest thermal conductivity of 22W/cmK.  Meanwhile, when we turn to its electrical properties, diamond is a rather featureless electrical insulator. However, with boron doping, it becomes a p-type semiconductor, with boron acting as a charge acceptor (*1-2*).  It is a promising material for electrical applications (*3*) such as high frequency and high power devices (*4*) owing to its high breakdown field (>10MV/cm) and high carrier mobility.

On the other hand, a heavily boron-doped diamond shows metallic conduction and it has been in use as electrodes in the field of electrochemistry (*5-6*).  Its physical properties however, have remained largely unexplored, particularly at low temperatures. Therefore the recent news of superconductivity in heavily boron-doped diamond synthesized by high pressure sintering was received with considerable surprise (*7*). Opening up new possibilities for diamond-based electrical devices, a systematic investigation of these phenomena clearly needs to be achieved.

Application of diamond to actual devices requires it to be made into the form of wafers or thin films.  The only procedures at present available to this end are low-pressure synthesis methods such as the Chemical Vapor Deposition (CVD) (*8-9*). In this letter we present unambiguous evidence of superconductivity in a diamond thin film deposited by a CVD method.  The onset of superconducting transition is found to be 7.4K, which is higher than the reported value in ref(7) and well above liquid helium temperature.  This finding, as discussed below establishes the superconductivity to be a universal property of boron-doped diamond, demonstrating that device application is indeed a feasible challenge.

A principle advantage of diamond thin films deposited by the CVD method is that it can contain boron at relatively higher concentrations compared to the bulk diamonds synthesized at high pressure.  Especially in the (111) oriented thin film, boron can be doped at a rate of about one order higher than in (001) oriented samples (*10*).  To achieve superconductivity, it is apparently crucial to realize a carrier concentration sufficiently high to induce an insulator-to-metal transition.  We are thus lead to expect the (111) oriented thin film to be a strong candidate for a superconductor with transition temperature above liquid helium temperature or higher.

The heavily boron-doped polycrystalline diamond thin film was deposited on a silicon (001) substrate using microwave plasma assisted chemical vapor deposition (MPCVD) method (*9*). The silicon substrates were pretreated by ultrasonic wave using diamond powder. Deposition was carried out under the condition of 50 Torr chamber pressure, 500 W microwave power and 800-900°C substrate temperature using a dilute gas mixture of methane and trimethylboron (TMB) in hydrogen. Methane concentration was 1% in hydrogen with B/C ratio of 2500 ppm. After 9 hour-deposition, a film of 3.5 μm thickness was obtained.

Scanning electron microscopy (SEM) image of the film is shown in Fig 1. The film morphology reveals to consist predominantly of {111} facets with a mean grain size of 1 μm. The X-ray diffraction pattern was obtained with Cu Kα radiation (wavelength = 0.154nm). A sharp peak was detected at 2θ = 43.9 degree corresponding to the (111) refraction of cubic cell of diamond structure. Scarce detections of (220), (311) and (400) peaks suggest the {111} textured growth of the film.

The transport properties were measured between room temperature and 1.7K. Fig. 2 shows the temperature dependence of resistivity of the film under several values of magnetic fields up to 9T. With decreasing temperature, the resistivity initially decreases slightly but increases gradually below 200K. The resistivity began to drop at around 7.4K which corresponds to the onset of a superconducting transition, and dropped to zero at around 4.2K ($T_C$ offset) in the absence of the field. The superconducting transition temperature is shifted with the increasing of the applied field. The field dependence of the onsets and offsets of $T_C$ are plotted in Fig. 2b. The extrapolation of $T_C$ onset approaches the value of 10.4 T. Assuming the dirty limit, the upper critical field $H_{C2}$ is estimated to be 7 T. This value is roughly similar to the $H_{C2}^{//c}$ of c-axis direction in $MgB_2$ (*11,12*). The irreversibility field is found to be 5.12 T at 0K. We have also confirmed reproducibility of superconductivity in different samples, the systematics of which shall be reported elsewhere.

The magnetization properties were measured by a superconducting quantum interface device (SQUID) magnetometer down to 1.78K. The temperature dependence of the magnetization is plotted in Fig 3a. Diamagnetic signals corresponding to superconductivity appeared below 4K where the resistance drops to zero. The large difference between the zero-field cooling (ZFC) and field cooling (FC) curvatures indicates that the material has a fairly large flux pinning force resulting in the trapping of magnetic flux in the field cooling condition.

Magnetization versus magnetic field (M-H) curvature obtained at 1.8K was plotted in fig 3b. Large symmetric hysteresis curvature shows the characteristics of typical type II superconductors. From the hysteresis of the magnetization curve, $\Delta M$(emu/cm$^3$), we can estimate critical current density $J_C$ on the assumption of a critical-state model with the simple formula, $J_C = 30 x \Delta M /d$, where d is the size of the sample (*13*). The $J_C$ at 0T is estimated to be 200 A/cm$^2$. The hardest superconducting wires will be fabricated.

Having established firmly the uniformity of superconductivity in these materials, we finally turn briefly to physical implications. Theoretical proposals of ref(14-16) have very recently been put forth which make the case that boron-doped diamond maybe viewed as a 3d analogue of $MgB_2$, and could be understood qualitatively in terms of a phonon mechanism incorporating the McMillan relation. Based on Hall conductivity measurements, we estimate the carrier concentration of our sample to be 9.4 x10$^{20}$/cm$^3$, which corresponds to a boron doping rate of 0.53%. In fact we have confirmed superconductivity in samples with doping rates as low as 0.18%. These values are considerably lower than those reported for materials sintered at high pressure, as well as theoretical estimates based on a first-principle evaluation incorporating phonon dynamics. These discrepancy may suggest (a) higher efficiency of doping in our samples, and (b) a stronger electron-phonon coupling than previously anticipated. A moderate Coulomb repulsion maybe another factor which needs to be considered in future theoretical treatments.

**Figures**

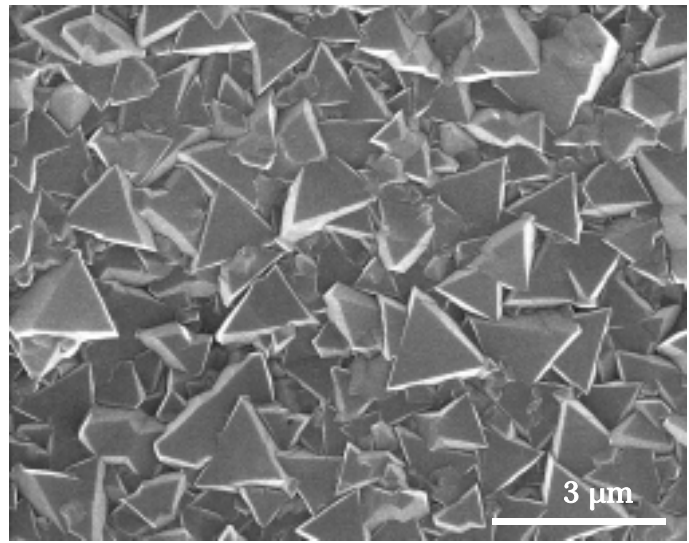

**Fig. 1.** Scanning electron microscopy (SEM) image of the film deposited by microwave plasma assisted chemical vapor deposition (MPCVD) method. The film morphology consist predominantly of {111} facets.

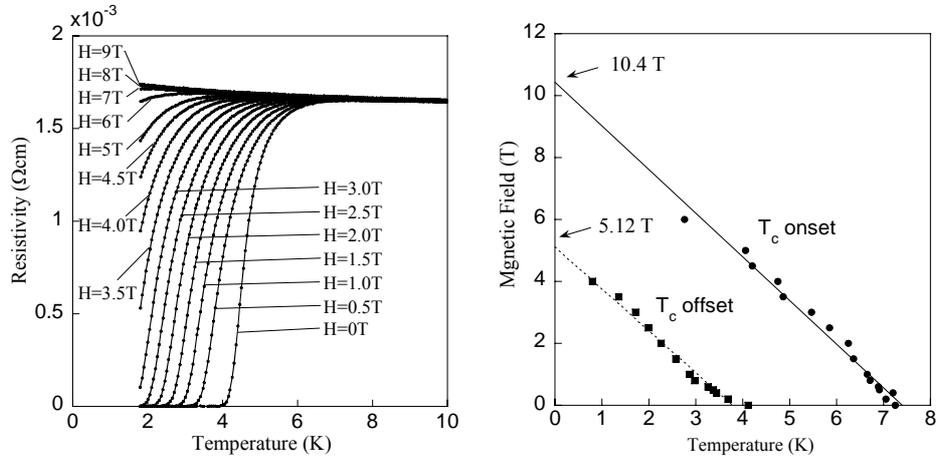

**Fig. 2.** (a) Temperature dependence of resistivity under several values of magnetic fields. In the absence of the field, the resistivity began to drop at around 7K which corresponds to $T_C$ onset, and dropped to zero at around 4.2K ($T_C$ offset). (b) The field dependence of the onsets and offsets of $T_C$. The $H_{C2}$ and irreversibility field are estimated to be 7 and 5.12 T, respectively.

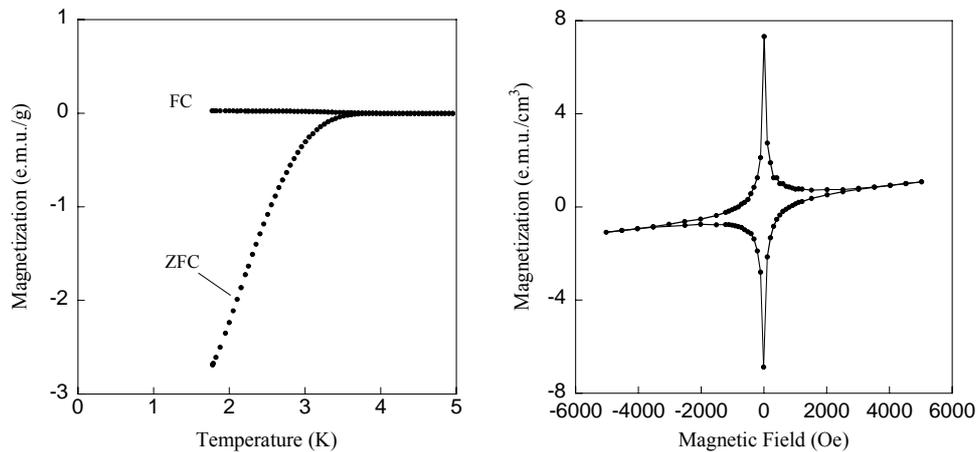

**Fig. 3.** (a) Temperature dependence of the magnetization at magnetic field of 1 Oe measured under the zero-field cooling (ZFC) and field cooling (FC) conditions. (b) Magnetization versus magnetic field (M-H) curvature obtained at 1.8K.